\documentclass[aps,preprint,epsfig,rotate]{revtex4}
\usepackage{graphicx}
\usepackage{bm}
\usepackage{epsfig}

\begin{document}
\title{Density functional theory and non-relativistic photoelectric effects in the few-electron atomic systems}

\author{Alexei M. Frolov}
 \email[E--mail address: ]{alex1975frol@gmail.com}  

\affiliation{Department of Applied Mathematics \\
 University of Western Ontario, London, Ontario N6H 5B7, Canada}

\date{\today}

\begin{abstract}

Closed analytical formulas are derived for the differential and total cross sections of the non-relativistic 
photoelectric effect in the three main classes of few-electron atomic systems: (1) neutral atoms and positively 
charged atomic ions which contain more than one bound electron, (2) negatively charged atomic ions, and (3) 
one-electron atoms and ions. Our procedure developed in this study is a combination of QED methods and results 
of the density functional theory obtained for atoms and ions. In all these systems the photoelectric effect is 
considered as photodetachment of the outer-most electron and our analysis is based on the results of density 
functional theory obtained for the electron density (radial) distribution in these atomic systems. Analytical 
formulas (similar to ours) for the differential and total cross sections of photoelectric effect for atomic 
systems from classes (1) and (2) have never been produced in earlier studies. 

\end{abstract}

\maketitle


\section{Introduction}

In general, photoelectric effect of atomic system is defined as absorption of radiation (or photon) by an atom 
in one of its bound states accompanied by the injection of one of the atomic electrons into a state of unbound 
. In other words, the bound `initial' atomic electron makes a transition of into the final unbound state. This 
unbound (or 'free') electron moves away from the parental atom/ion in the Coulomb field of the final, positively 
charged atomic ion. In the case of negatively charged ions such an unbound photo-electron moves in the field of 
a neutral atom. Theory of the non-relativistic photoelectric effect (everywhere below, photoeffect, for short), 
or theory of photoionization of light atoms and ions is a well developed chapter of Quantum Electrodynamics, or 
QED, for short (see, e.g., \cite{AB}, \cite{BLP}). In particular, the closed formula for the photoionization 
cross section of one-electron atoms and ions was produced by Stobbe \cite{Stob}, but it was restricted to the 
case of ground states in these atomic systems. Since 1930's Stobbe's formula was extensively used to explain and 
describe many actual processes in various physical systems and devices where atomic photoionization plays a 
crucial role. Twenty years later analogous formula has been derived for the photoionization cross sections from 
the excited states in one-electron atoms and ions (see, discussion in \cite{BS} and references therein). However, 
nobody (even Hans Bethe) could derive the closed analytical formulas which can accurately describe similar 
photoionization cross sections of the few-electron atomic systems. On the other hand, Chandrasekhar and others 
were unsuccessfully trying for a number of years to derive the closed analytical formula for the photodetachment 
cross section of the negatively charged, hydrogen ion H$^{-}$. 

In this communication we are happy to report that these long-standing `unsolvable' problems of atomic photoionization
(and photodetachment) have finally been solved and here we describe the corresponding analytical solutions. Our main 
goal in this study is to obtain the closed, analytical formulas for the both differential $\frac{d \sigma}{d o}$ and 
total $\sigma$ cross sections of photoionization and/or photodetachment of the few- and many-electron atoms and ions. 
All these cross sections must be derived as the explicit functions of the atomic ionization potential $I(Q, N_e)$ and 
$\omega$ is the cyclic frequency of the incident light. The atomic ionization potential $I(Q, N_e)$ is the function of 
the total number of bounded electrons $N_e$ in the initial atom/ion and electrical charge $Q$ of the atomic nucleus. 
To achieve this goal we investigate photoeffect for the outer-most atomic electron, or in other words, photodetachment 
of the outer-most electron(s) in atomic systems which include few- and many-electron atoms and ions. Below, we shall 
not consider photoionization of electrons from internal electron shells, double photoionization, photodetachment with 
instantaneous excitations of other electron transitions and other similar processes. Instead, we will focus on the 
usual, or thermal photoionization of that electron which is least weakly bound to the central nucleus. Investigation 
of such processes is reduced to the analysis of photodetachment of the outer-most electron(s) in atomic system. It is 
clear that the photodetachment cross sections of the outer-most electron(s) in few-electron atomic systems coincide
with the corresponding cross sections of photoionization and/or photodetachment of the whole atom/ion. Briefly, by 
considering photodetachment of the outer-most electrons in different atomic systems we can accurately describe the 
regular (or thermal) photoionization and photodetachment in arbitrary few- and many-electron atomic systems. Currently, 
there is some misconception (see, e.g., discussion in \cite{Star} and references therein) that analytical methods for 
dealing with atomic photoionization can be applied only to one-electron atoms and ions. Below, we show that this 
statement is wrong and many photoionization and photodetachment cross sections can be approximated to very good 
accuracy by using old-fashion and relatively simple analytical methods and better understanding of the physics of 
photoeffect. Our working method in this study is a combination of QED approach and results of the density functional 
theory. 

\section{Differential cross section and wave functions}

According to the rules of Quantum Electrodynamics (see, e.g., \cite{AB}, \cite{BLP}) the differential cross section 
of the non-relativistic photoeffect for an arbitrary atomic system is written in the following form \cite{BLP}  
\begin{eqnarray}
 d\sigma = \frac{e^{2} m \mid {\bf p} \mid}{2 \pi \omega} \mid {\bf e} \cdot {\bf v}_{fi} \mid^{2} do = \frac{p}{2 
 \pi \omega a_0} \mid {\bf e} \cdot {\bf v}_{fi} \mid^{2} do \; \; \; \label{cross}
\end{eqnarray}
where $m \approx$ 9.1093837015$\cdot 10^{-28}$ $g$ is the rest mass of electron, while $- e$ is its electric charge. 
Also in this equation $a_0 \approx$ 5.29177210903 $\cdot 10^{-9}$ $cm$ is the Bohr radius and ${\bf e}$ is the vector 
which describes the actual polarization of initial photon. Also, in this formula $p = \mid {\bf p} \mid$ is the 
momentum of the final (or free) photo-electron, $\omega$ is the cyclic frequency of the incident light quanta, while 
${\bf v}_{fi}$ is the matrix element of the `transition' velocity ${\bf v}_{fi}$ between initial $| i \rangle$ and 
final $\langle f |$ states. For this matrix element we can write ${\bf v}_{fi} = - \frac{\imath}{m} \langle \psi_{f} 
| \nabla | \psi_{i} \rangle$, where the notation $\psi$ designates the wave functions, while the indexes $f$ and $i$ 
in this equation and in all formulas below stand for the final and initial states, respectively. The formula, 
Eq.(\ref{cross}), is written in the relativistic units, where $\hbar = 1, c = 1, e^{2} = \alpha$ and $\alpha \approx$ 
7.2973525693$\cdot 10^{-3}$ ($\approx \frac{1}{137}$) is the dimensionless fine-structure constant. These relativistic 
units are convenient to perform analytical calculations in Quantum Electrodynamics (below QED, for short). However, in 
order to determine the non-relativistic cross-sections it is better to apply either the usual units, e.g., $C G S$ 
units, or atomic units in which $\hbar = 1, m_e = 1$ and $e = 1$. In atomic units the formula, Eq.(\ref{cross}), takes 
the form
\begin{eqnarray}
 d\sigma = \frac{\alpha a^{2}_{0} p}{2 \pi \omega} \mid {\bf e} \cdot \langle \psi_{f} | \nabla | \psi_{i} \rangle 
 \mid^{2} do \; \; , \; \; \label{crossau}
\end{eqnarray}
where $a_{0} = \frac{\hbar^{2}}{m e^{2}}$ is the Bohr radius and $\alpha = \frac{e^{2}}{\hbar c}$ is the dimensionless 
fine-structure constant. In atomic units we have $a_{0} = 1$ and $\alpha = \frac{1}{c} (\approx \frac{1}{137})$. Let 
us assume that initial electron was bound to some atomic system, i.e., to a neutral atom, negatively and/or positively 
charged ion. The energy of this bound state (or discrete level) is $\varepsilon = - I$, where $I$ is the atomic 
ionization potential. It is clear that the condition $\omega \ge I$ which must be obeyed to make photoeffect possible. 
In fact, for atomic photoeffect we always have the following relation $\omega = I + \frac12 p^{2}$ between $\omega, I$ 
and $p$, where $p$ is the momentum of the final photo-electron. This equation is also written in the form $p = \sqrt{2 
(\omega - I)}$.  
  
\subsection{Wave functions of the final and initial states}
 
Now, we need to develop some logically closed procedure to calculate the matrix element (or transition amplitude) $\langle 
\psi_{f} | \nabla | \psi_{i} \rangle$ which is included in the formulas, Eqs.(\ref{cross}) and (\ref{crossau}). Everywhere 
below in this study, we shall assume that the original (or incident) atomic system was in its lowest-energy ground (bound) 
state which is usually $S(L - 0)-$state. For one-electron atomic system this state is always the doublet $1^{2}s(\ell = 
0)-$state. Then, in the lowest order dipole approximation \cite{AB} the outgoing (or final) photo-electron will move in the 
$p(\ell = 1)-$wave. By using this fact we can write the wave function of the final electron
\begin{eqnarray}
 \psi_{f}({\bf r}) = \frac{2 \ell + 1}{2 p} \; \; P_{\ell}({\bf n} {\bf n}_1) \; \psi_{\ell;p}(r) = ({\bf n} {\bf n}_1) 
 \; \; \frac{3}{2 p} \psi_{1;p}(r) = ({\bf n} {\bf n}_1) R_{1;p}(r) \; , \; \label{psi-f}
\end{eqnarray} 
where $\ell = 1, P_{\ell}(x)$ is the Legendre polynomial (see, e.g., \cite{GR}) and $R_{\ell=1;p}(r) = R_{1;p}(r)$ is 
the corresponding radial function which depends upon the explicit form of the interaction potential between outgoing 
photo-electron and remaining atomic system. Also, in this formula the unit vectors ${\bf n}$ and ${\bf n}_1$ are: ${\bf 
n} = \frac{{\bf p}}{p}$ and ${\bf n}_1 = \frac{{\bf r}}{r}$. The unit vector ${\bf n}$ determines the direction of 
outgoing (or final) photo-electron, or ${\bf p}-$direction, for short. 

If the interaction potential between outgoing photo-electron and remaining atomic system is described by a Coulomb 
potential, then the radial function $\psi_{\ell;p}(r)$ in Eq.(\ref{psi-f}) is the normalized Coulomb function of the 
first kind (see, e.g., \cite{AS}) which is 
\begin{eqnarray}
 \psi_{\ell;p}(r) = \frac{2^{\ell} Z}{(2 \ell + 1)!} \sqrt{\frac{8 \pi}{\nu [1 - \exp(-2 \pi \nu)]}} \; \; \; 
 \prod^{\ell}_{s=1} \sqrt{s^{2} + \nu^{2}} \; \; (p r)^{\ell} \; \exp(-\imath p r) \; \times \nonumber \\
 \; {}_1F_{1}(\ell + 1 + \imath \nu, 2 \ell + 2; 2 \imath p r) \; , \; \label{radiala0}
\end{eqnarray} 
where ${}_1F_{1}(a, b; z)$ is the confluent hypergeometric function defined exactly as in \cite{GR} and \cite{AS}, 
$\nu = \frac{Z}{p} = \frac{Q - N_e + 1}{p}$ and $Z = Q - N_e + 1$ is the electric charge of the final atomic 
fragment. From this equation one finds the wave function of an electron which moves in a central Coulomb field in 
the $p-$wave (i.e., for $\ell = 1$) 
\begin{eqnarray}
 \psi_{1;p}(r) &=& \frac{2 Z p}{3!} \sqrt{\frac{8 \pi (1 + \nu^{2})}{\nu [1 - \exp(-2 \pi \nu)]}} \; \; r 
 \exp(-\imath p r) \; \; {}_1F_{1}(2 + \imath \nu, 4; 2 \imath p r) \; . \; \label{radiala}
\end{eqnarray} 
By multiplying this formula by the additional factor $\frac{3}{2 p}$ from Eq.(\ref{psi-f}), we can write for this 
wave function with $\ell = 1$ (in atomic units) 
\begin{eqnarray}
 R_{1;p}(r) &=& Z \sqrt{\frac{2 \pi (1 + \nu^{2})}{\nu \Bigl(1 - \exp(-2 \pi \nu)\Bigr)}} \; \; r \exp(-\imath 
 p r) \; \; {}_1F_{1}(2 + \imath \nu; 4; 2 \imath p r) \; \; \nonumber \\
  &=& p \sqrt{\frac{2 \pi \nu (1 + \nu^{2})}{1 - \exp(-2 \pi \nu)}} \; \; r \exp(-\imath p r) \; \; 
 {}_1F_{1}(2 + \imath \nu; 4; 2 \imath p r) \; , \; \label{rad1}
\end{eqnarray} 
where the electric charge $Z$ of the remaining atomic system is an increasing function of the nuclear charge $Q$, 
but it also depends upon the total number of bound electrons $N_e$. All phases and normalization factors in this 
formula coincide exactly with their values presented in \cite{LLQ}. 

For the non-Coulomb (or short-range) interaction potentials between outgoing photo-electron and remaining atomic 
cluster the normalized radial wave function of the continuous spectra is written as a product of the spherical 
Bessel function $j_{\ell}(p r)$ and a factor which equals $2 p$, i.e., $\psi_{\ell;p}(r) = 2 p j_{\ell}(p r)$ 
(see, e.g., \cite{LLQ}, \$ 33). For $\ell = 1$ one finds $\psi_{1;p}(r) = 2 p j_{1}(p r)$. This means that the 
function $\psi_{\ell=1;p}(r)$ is regular at $r = 0$. From here for the radial function $R_{1;p}(r) = \frac{3}{2 
p} \psi_{\ell=1;p}(r)$ we obtain 
\begin{eqnarray}
 R_{1;p}(r) = \frac{3}{2 p} \Bigl[ 2 p j_{\ell=1}(p r) \Bigr] = 3 j_{\ell=1}(p r) = 3 \sqrt{\frac{\pi}{2 p r}} 
 \; \; J_{\frac32}(p r) \; , \; \label{rad2}
\end{eqnarray}
where $j_{1}(x) = \frac{sin x}{x^{2}} - \frac{cos x}{x}$ and $J_{\frac32}(x)$ is the Bessel function which is 
regular at the origin (at $r = 0$) and defined exactly as in \cite{GR}. These two radial wave functions of an 
unbound photo-electron, Eqs.(\ref{rad1}) and (\ref{rad2}), are used in our calculations below. 
 
Now, let us discuss the wave functions of the initial atomic system which has one nucleus with the electric charge 
$Q$ and $N_e$ bound electrons. By analyzing the current experimental data it is easy to understand that the 
non-relativistic photodetachment of the outer-most electrons in a few- and many-electron atomic systems is produced 
by photons with large and very large wavelengths $\lambda$. In reality, the wavelengths $\lambda$ of incident light 
quanta substantially exceed the actual sizes of atoms and ions which are included in this process. For instance, the 
wavelengths $\lambda$ of photons that produce photodetachment of the negatively charged hydrogen ions H$^{-}$ in 
Solar photosphere exceed 7000 $\AA \approx$ 13232 $a.u.$, while the spatial radius $R$ of this ion equals $\approx$ 
2.710 $a_0$ = 2.710 $a.u.$, i.e., $R \ll \lambda$. In fact, for the H$^{-}$ ion its spatial radius $R$ is smaller 
than the wavelengths $\lambda$ of incident light quanta in thousands times. In other words, photodetachment of the 
outer-most electron in the H$^{-}$ ion is produced at large and very large distances from the central atomic nucleus 
and from the second atomic electron. Such asymptotic spatial areas in the H$^{-}$ ion are very important to determine 
photodetachment cross sections, since only in these spatial areas one finds a relatively large overlap between the 
electron and photon wave functions. 

Similar situations can be found in other atomic systems considered in this study, e.g., for all neutral atoms and 
positively charged ions. In each of these Coulomb systems photodetachment of the outer-most electron(s) mainly 
occurs in the asymptotic areas of their wave functions. These asymptotic areas are located far and very far form 
the central atomic nucleus and other internal atomic electrons. Therefore, in our analysis of non-relativistic 
photodetachment of the outer-most electron(s) we can restrict ourselves to large spatial areas and consider only 
the long-range asymptotic of these wave functions. Moreover, it seems very tempting to neglect the small area of 
electron-electron correlations around the central nucleus ($R \le a_0$) and consider the long-range asymptotics of 
atomic wave functions as the `new' wave functions for our problem. Briefly, in this procedure we replace the actual 
wave functions for each of these atomic systems by their long-range radial asymptotics. It is clear that the `new' 
bound state wave function is one-electron function and it has a different normalization constant. Obviously, this 
is an approximation, but as follows from our results the overall accuracy of our approximation is very good and 
sufficient to describe photodetachment of the outer most-electrons in all atomic systems discussed in this study. 

In this study we consider the photoeffect in the three different classes of atomic systems: (a) atom/ion which 
initially contains $N_e$ bound electrons, while its nuclear charge $Q$ ($Q \ge N_e$) is arbitrary, (b) one-electron 
atom/ion, where $N_e = 1$ and $Q$ is arbitrary, and (c) negatively charged ion where $Q = N_e - 1$ and the both $Q$ 
and $N_e$ are arbitrary. As is well known (see, e.g., \cite{Osten} and references therein) in arbitrary atomic $(Q, 
N_e)$-system the radial wave function of the ground $S(L = 0)$-states has the following long-distance (radial) 
asymptotics 
\begin{eqnarray}
 R_{i}(r) &=& C(b;I) r^{b - 1} \exp(- \sqrt{2 I} r) = \frac{(2 \sqrt{2 I})^{b + \frac12}}{2 \sqrt{\pi 
 \Gamma(2 b + 1)}} \; \; r^{b - 1} \; \exp(- \sqrt{2 I} r) \; \nonumber \\
 &=& \frac{(2 \sqrt{2 I})^{b + \frac12}}{2 \sqrt{\pi \Gamma(2 b + 1)}} \; \; r^{\frac{Q - N_e + 1}{\sqrt{2 
 I}} - 1} \; \exp(- \sqrt{2 I} r) \; \; \label{rad3}   
\end{eqnarray}
where $b = \frac{Q - N_e + 1}{\sqrt{2 I}} = \frac{Z}{\sqrt{2 I}}, I (\ge 0)$ is the atomic ionization potential 
and $Z = Q - N_e + 1$ is the electric charge of the remaining atom/ion. In Eq.(\ref{rad3}) and everywhere below 
the notation $\Gamma(x)$ denotes the Euler's gamma-function $\Gamma(1 + x) = x \Gamma(x)$, which is often called 
the Euler's integral of the second kind \cite{GR}. This important result of the Density Functional Theory (or 
DFT, for short) plays a crucial role in this study. Here we have to emphasize the following fundamental fact: the 
formula, Eq.(\ref{rad3}), is the exact long-range asymptotic of the truly correlated, $N_e$-electron wave function 
of an actual atom/ion. The derivation of this formula is not based on any approximation. In other words, by 
choosing the wave function $R_{i}(r)$ in the form of Eq.(\ref{rad3}) we do not neglect any of the electron-electron 
correlations in atomic wave function. 

However, since photodetachment of the outer-most electrons mainly occurs at large and very large distances from the 
atomic nucleus, then it will be a very good approximation to describe this phenomenon, if we continue the radial 
$R_{i}(r)$ function, Eq.(\ref{rad3}), on the whole real $r-$axis, including the radial origin, i.e., the point $r = 
0$. This allows us to determine the factor $C(b;I)$ in Eq.(\ref{rad3}) which is the normalization constant of the 
radial $R_{i}(r)$ function, which now continues on the whole real $r-$axis, including the radial origin, i.e., the 
point $r = 0$. Namely, after this step our analysis becomes approximate. Nevertheless, we can determine the 
normalization constant $C(b;I)$ for this radial $R_{i}(r)$ function, Eq.(\ref{rad3}) where $0 \le r < +\infty$. It 
equals  
\begin{eqnarray}
   C(b;I) = \frac{(2 \sqrt{2 I})^{b+\frac12}}{2 \sqrt{\pi \Gamma(2 b + 1)}} \; \; \label{norm}
\end{eqnarray}
where the atomic ionization potential $I$ and parameter $b$ are the two real, non-negative numbers. In some equations 
below the $\sqrt{2 I}$ value is also designated as $B$. In the general case, the atomic ionization potential $I$ is 
an unknown function of $Q$ and $N_e$. 

For the negatively charged (atomic) ions we always have $b = \frac{Q - N_e + 1}{\sqrt{2 I}} = 0$. The long-distance 
asymptotic of the radial wave function of an arbitrary negatively charged ion is always written in the form: $R(r) 
\sim \frac{C}{r} \exp(-\sqrt{2 I} r)$, where $C = \sqrt[4]{\frac{I}{2 \pi^2}}$ is the normalization constant and $I$ 
is an unknown function of $Q$ and $N_e$. In contrast with this, for one-electron atoms and ions we have $N_e = 1$ and 
$2 I = Q^{2}$ and ionization potential $I$ is the uniform function of the nuclear charge $Q$ only. For one-electron 
atomic systems we also have $b = 1$ and the exact wave function is written in the form $R(r; Q) = A \exp(- \sqrt{2 I} 
r)$, where $2 I = Q^{2}$, and $A = \frac{Q \sqrt{Q}}{\sqrt{\pi}}$ is the normalization constant. 

\subsection{Gradient operator and its matrix elements}

Let us derive some useful formulas for the matrix element $\langle \psi_{f} | \nabla | \psi_{i} \rangle$ which 
is included in Eq.(\ref{crossau}) and plays a central role in this study. It is clear that we need to determine 
the vector-derivative (or gradient) of the initial wave function, which is a scalar function. In general, for 
the interparticle (or relative) vector ${\bf r}_{ij} = {\bf r}_{j} - {\bf r}_{i}$ the corresponding gradient 
operator in spherical coordinates takes the form (see, e.g., \cite{Kochin})
\begin{eqnarray}
 \nabla_{ij} = \frac{d }{d {\bf r}_{ij}} = \frac{{\bf r}_{ij}}{r_{ij}} \frac{\partial }{\partial r_{ij}} + 
 \frac{1}{r_{ij}} \nabla_{ij}(\Omega) = {\bf e}_{r;ij} \frac{\partial }{\partial r_{ij}} + {\bf e}_{\theta;ij} 
 \frac{1}{r_{ij}} \frac{\partial }{\partial \theta_{ij}} + {\bf e}_{\phi;ij} \frac{1}{r_{ij} \sin\theta_{ij}} 
 \frac{\partial }{\partial \phi_{ij}} \; , \; \label{grad}
\end{eqnarray}
where $\nabla_{ij}(\Omega)$ is the angular part of the gradient vector which depends upon angular variables 
($\theta$ and $\phi$) only, while ${\bf e}_{r;ij} = \frac{{\bf r}_{ij}}{r_{ij}} = {\bf n}_{ij}, {\bf 
e}_{\theta;ij}$ and ${\bf e}_{\phi;ij}$ are the three unit vectors in spherical coordinates which are defined 
by the ${\bf r}_{j}$ and ${\bf r}_{i}$ vectors, where ${\bf r}_{j} \ne {\bf r}_{i}$. 

If the radial part of the initial wave function depends upon the scalar radial variable only, then all 
derivatives in respect to the both angular variables $\theta$ and $\phi$ equal zero identically and we can 
write    
\begin{eqnarray}
 \nabla_{ij} R(r) = \frac{{\bf r}_{ij}}{r_{ij}} \frac{\partial R(r_{ij})}{{\partial r}_{ij}} = \frac{{\bf 
 r}_{ij}}{r_{ij}} \frac{d R(r_{ij})}{d r_{ij}} = {\bf n}_{ij} \frac{d R(r_{ij})}{d r_{ij}} \; , \; \label{grad1}
\end{eqnarray} 
where ${\bf n}_{ij}$ is the unit vector in the direction of interparticle ${\bf r}_{ij}$ variable. For one-center 
atomic systems we can determine $r_{1j} = r_{j}$, and for one-electron systems $r_{12} = r_1 = r$. In this case, 
the formula, Eq.(\ref{grad1}), is written in the form: $\nabla R(r) = {\bf n}_1 \frac{d R(r)}{d r}$, where ${\bf 
n}_1 = \frac{{\bf r}_{1}}{r_{1}}$. In this notation the radial matrix element is 
\begin{eqnarray}
 & & {\bf e} \langle \psi_{f} | \nabla | \psi_{i} \rangle = \int_{0}^{+\infty} \Bigl\{\oint ({\bf n} \cdot {\bf 
 n}_1) ({\bf e} \cdot {\bf n}_1) do_1 \Bigr\} \Bigr( R_{1;p}(r) \frac{d R_{i}(r)}{d r} \Bigl) r^{2} dr \nonumber \\
 &=& \frac{4 \pi}{3} ({\bf e} \cdot {\bf n}) \int_{0}^{+\infty} \Bigr( R_{1;p}(r)\frac{d R_{i}(r)}{d r} \Bigl) 
 r^{2} dr = \frac{4 \pi}{3} ({\bf e} \cdot {\bf n}) I_{rd} \; , \; \label{ME}
\end{eqnarray} 
where ${\bf n}$ is the unit vector which determines the direction of propagation of the final electron, while 
$R_{1;p}(r)$ and $R_{i}(r)$ are the radial functions of the final and initial states, respectively. The notation 
$I_{rd}$ in this formula, Eq.(\ref{ME}), stands for the following auxiliary radial integral
\begin{eqnarray} 
 I_{rd} = \int_{0}^{+\infty} \Bigr( R_{1;p}(r) \frac{d R_{i}(r)}{d r} \Bigl) r^{2} dr = - \int_{0}^{+\infty} 
 \Bigr( R_{i}(r) \frac{d R_{1;p}(r)}{d r} \Bigl) r^{2} dr \; , \; \label{aux}
\end{eqnarray} 
where we used the so-called 'transfer of the derivative' (or partial integration) which often helps to simplify 
analytical calculations of this radial integral. 

By substituting the expression, Eq.(\ref{ME}), into the formula, Eq.(\ref{crossau}), one finds the following 
'final' formula for the differential cross section of the non-relativistic photodetachment of an arbitrary 
atomic system
\begin{eqnarray}
 d\sigma = \frac{16 \pi^{2}\alpha a^{2}_{0} p}{18 \pi \omega} ({\bf e} \cdot {\bf n})^{2} \mid I_{rd} 
 \mid^{2} do = \frac{8 \pi\alpha a^{2}_{0} p}{9 \omega} ({\bf e} \cdot {\bf n})^{2} \mid I_{rd} \mid^{2} do 
 \; . \; \label{sigma-a}
\end{eqnarray} 
As follows from this formula the angular distribution of photo-electrons is determined by the 'angular' factor 
$({\bf e} \cdot {\bf n})^{2}$. This cross section of photodetachment corresponds to the truly (or 100 \%) 
polarized light. However, in many actual applications the incident beam of photons is unpolarized and we deal 
with the natural (or white) light. If the incident beam of photons was unpolarized, then we need to apply the 
formula $\overline{({\bf e} \cdot {\bf n})^{2}} = \frac12 ({\bf n}_l \times {\bf n})^{2}$, where ${\bf n}_l$ 
is the unit vector which describes the direction of incident light propagation and ${\bf n}$ is the unit vector 
which determines the direction of propagation of the final photo-electron. In this study the notation $({\bf a} 
\times {\bf b})$ denotes the vector product of the two vectors ${\bf a}$ and ${\bf b}$. Finally, the differential 
cross section of photodetachment is written in the form 
\begin{eqnarray}
 d\sigma = \Bigl(\frac{4 \pi \alpha a^{2}_{0} p}{9 \omega}\Bigr) \; ({\bf n}_{l} \times {\bf n})^{2} \mid 
 I_{rd} \mid^{2} do = \frac{4 \pi}{9} \alpha a^{2}_{0} \;  \Bigl(\frac{p}{\omega}\Bigr) \; \sin^{2}\Theta 
 \mid I_{rd} \mid^{2} do \; , \; \label{sigma-aa}
\end{eqnarray} 
where $\Theta$ is the angle between two unit vectors ${\bf n}_{l}$ and ${\bf n}$. The presence of vector 
product $({\bf n}_{l} \times {\bf n})^{2}$ in Eq.(\ref{sigma-aa}) is typical for the dipole approximation. 
As follows from the formula, Eq.(\ref{sigma-aa}), analytical and numerical calculations of the differential 
cross section of photodetachment are now reduced to analytical computations of the auxiliary radial integral 
$I_{rd}$, Eq.(\ref{aux}). The total cross section of photodetachment is 
\begin{eqnarray}
 \sigma = \Bigl(\frac{32 \pi^{2} \alpha a^{2}_{0} p}{27 \omega}\Bigr) \; \mid I_{rd} \mid^{2} = \frac{32 
 \pi^{2} \alpha a^{2}_{0}}{27} \; \Bigl(\frac{p}{\omega}\Bigr) \mid I_{rd} \mid^{2} \; . \; \label{sigma-aa1}
\end{eqnarray} 
By using different expressions for the initial and final wave functions we can determine the differential 
and total cross sections of photodetachment of the outer most electrons in various few- and many-electron 
atoms and ions. The corresponding formulas are presented below. 

\section{Few-electron neutral atoms and positively charged ions} 

First, let us consider photodetachment of the outer-most electron(s) in few-electron neutral atoms, where 
$Q = N_e$ and $N_e \ge 2$, and in positively charged atomic ions, where $Q > N_e$ and $N_e \ge 2$. In both 
these cases the final sub-systems, i.e., outgoing photo-electron and remaining positively charged ion, 
interact with each other by an attractive Coulomb potential. For atoms and positively charged ions this 
process obviously coincides with atomic photoionization. The wave function of outgoing photo-electron must 
be taken in the form of Eq.(\ref{rad1}), while the wave function of the initial atomic state is chosen in 
the form of Eq.(\ref{rad3}). The radial derivative of this initial wave function is
\begin{eqnarray} 
  \frac{d}{d r} \Bigl[ r^{b - 1} \exp(- B r) \Bigr] = (b - 1) r^{b - 2} \exp(- B r) - B r^{b - 1} 
 \exp(- B r) \; \; , \; \label{diff1} 
\end{eqnarray}  
where $b = \frac{Q - N_e + 1}{\sqrt{2 I}} = \frac{Z}{\sqrt{2 I}} = \frac{Z}{B}, Z = Q - N_e + 1$ and $B = 
\sqrt{2 I}$. Therefore, the formula for our auxiliary radial integral $I_{rd}$, Eq.(\ref{aux}), includes 
the two terms $I_{rd} = I^{(1)}_{rd} + I^{(2)}_{rd}$, where 
\begin{eqnarray} 
 & &I^{(1)}_{rd} = p \sqrt{\frac{2 \pi \nu (1 + \nu^{2})}{1 - \exp(-2 \pi \nu)}} C(b;B) (b - 1)
 \int_{0}^{+\infty} r^{(b + 2) - 1} \exp(- B r -\imath p r) {}_1F_{1}(2 + \imath \nu; 4; \nonumber \\
 & &2 \imath p r) dr = p \sqrt{\frac{2 \pi \nu (1 + \nu^{2})}{1 - \exp(-2 \pi \nu)}} C(b;B) \; \frac{(b 
 - 1) \Gamma(b + 2)}{(B + \imath p)^{b + 2}} \; \; {}_2F_{1}(2 + \imath \nu; b + 2; 4; \frac{2 \imath p}{B 
 + \imath p} \Bigr) \; \; \label{rint1} \\
 &=&p \sqrt{\frac{2 \pi \nu (1 + \nu^{2})}{1 - \exp(-2 \pi \nu)}} C(b;B) \; \frac{(b - 1) \Gamma(b 
 + 2)}{(B + \imath p)^{b + 2}} \Bigl(\frac{B + \imath p}{B - \imath p}\Bigr)^{\imath \nu + b} \; \; 
 {}_2F_{1}(2 - \imath \nu; 2 - b; 4; \frac{B - \imath p}{B + \imath p} \Bigr) \; , \; \nonumber  
\end{eqnarray}  
where ${}_2F_{1}(a, b ; c; z)$ is the (2,1)-hypergeometric function defined exactly as in \cite{GR}, $\nu 
= \frac{Z}{p}$ and $Z = Q - N_e + 1$, while $C(b;B)$ is the normalization constant of the bound state 
radial function (see,  Eq.(\ref{norm})). The explicit formula for the second radial integral $I^{(2)}_{rd}$ 
is 
\begin{eqnarray} 
 &&I^{(2)}_{rd} = - p \sqrt{\frac{2 \pi \nu (1 + \nu^{2})}{1 - \exp(-2 \pi \nu)}} C(b;B) \; B 
 \int_{0}^{+\infty} r^{(b + 3) - 1} \exp(- B r -\imath p r) {}_1F_{1}(2 + \imath \nu; 4; \nonumber \\
 &&\imath 2 p r) dr = - p \sqrt{\frac{2 \pi \nu (1 + \nu^{2})}{1 - \exp(-2 \pi \nu)}} C(b;B) \; 
 \frac{B \; \Gamma(b + 3)}{(B + \imath p)^{b + 3}} \; \; {}_2F_{1}(2 + \imath \nu; b + 3; 4; 
 \frac{2 \imath p}{B + \imath p} \Bigr) \; \label{rint2} \\
 &&= -p \sqrt{\frac{2 \pi \nu (1 + \nu^{2})}{1 - \exp(-2 \pi \nu)}} C(b;B) \; \frac{B \; \Gamma(b 
 + 3)}{(B + \imath p)^{b + 3}} \Bigl(\frac{B + \imath p}{B - \imath p}\Bigr)^{\imath \nu + b + 1} 
 \; \; {}_2F_{1}(2 - \imath \nu, 1 - b; 4; \frac{B - \imath p}{B + \imath p} \Bigr) \; . \; \nonumber  
\end{eqnarray} 
where $C(b;B)$ is the normalization constant, Eq.(\ref{norm}), and $B = \sqrt{2 I}$, where $I$ is the 
ionization potential of the initial atomic systems. To simplify the two last formulas we note that 
\begin{eqnarray} 
 \Bigl(\frac{B + \imath p}{B - \imath p}\Bigr)^{\imath \nu} = \Bigl[ \Bigl(\frac{\frac{\nu}{b} + 
 \imath}{\frac{\nu}{b} - \imath}\Bigr)^{\imath \frac{\nu}{b}} \Bigr]^{b} = \exp\Bigl[- 2 \nu \; 
 {\rm arccot} \Bigl(\frac{\nu}{b}\Bigr)\Bigr] \; \label{artg}
\end{eqnarray} 
where $\nu = \frac{Z}{p} = \frac{Q - N_e + 1}{p}$ and $b = \frac{Q - N_e + 1}{\sqrt{2 I}}$ (here $Z = Q 
- N_e + 1$, see above) and $\frac{B}{p} = \frac{\nu}{b}$, or $\nu = \frac{b B}{p}$. The function ${\rm 
arccot} \; x$ is the inverse cotangent function which is also equals ${\rm arccot} \; x = \arccos 
(\frac{x}{\sqrt{1 + x^{2}}})$ (this formula is used in numerical calculations). 

After a few additional, relatively simple transformations we derive to the following expression for the 
total radial integral $I_{rd} = I^{(1)}_{rd} + I^{(2)}_{rd}$:
\begin{eqnarray} 
 &&I_{rd} = \frac{p^{2}}{b} \sqrt{\frac{\nu (1 + \nu^{2})}{1 - \exp(-2 \pi \nu)}} \; \frac{2^{b} B^{b} 
 \sqrt{B} \Gamma(b + 2) (B - \imath p)^{1 - b}}{\sqrt{\Gamma(2 b + 1)} (B^2 + p^{2})^{2}} \exp\Bigl[- 2 
 \nu \; {\rm arccot} \Bigl(\frac{\nu}{b}\Bigr)\Bigr] \Bigr[ (b - 1) \times \nonumber \\
 &&(\nu - \imath b) \; \; {}_2F_{1}(2 - \imath \nu; 2 - b; 4; \frac{\nu - \imath b}{\nu + \imath b} 
 \Bigr) - \nu \; (b + 2) \; \; {}_2F_{1}(2 - \imath \nu, 1 - b; 4; \frac{\nu - \imath b}{\nu + \imath b} 
 \Bigr) \Bigr]  \; . \; \label{Ird}
\end{eqnarray}
From this expression one easily finds the following formula for the $\mid I_{rd} \mid^{2}$ value 
\begin{eqnarray} 
 &&\mid I_{rd} \mid^{2} = \frac{4^{b} \; \nu^{2 b + 2} (1 + \nu^{2}) \Gamma^{2}(b + 2)}{p (\nu^{2} 
 + b^{2})^{b+3} (1 - \exp(-2 \pi \nu)) \Gamma(2 b + 1)} \; \exp\Bigl[- 4 \nu \; {\rm arccot} 
 \Bigl(\frac{\nu}{b}\Bigr)\Bigr] \; \Bigr| (b - 1) \times \nonumber \\
 && (\nu - \imath b) \; \; {}_2F_{1}(2 - \imath \nu; 2 - b; 4; \frac{\nu - \imath b}{\nu + \imath b} 
 \Bigr) - \nu \; (b + 2) \; {}_2F_{1}(2 - \imath \nu, 1 - b; 4; \frac{\nu - \imath b}{\nu + \imath b} 
 \Bigr) \Bigr|^{2} \; . \; \label{I2rdmod}
\end{eqnarray}
By multiplying this expression by the $\Bigl(\frac{8 \pi \alpha a^{2}_{0} p}{9 \omega}\Bigr) \; ({\bf n} 
\cdot {\bf e})^{2}$ factor one finds the final formula for the differential cross section of photoionization 
of few- and many-electron neutral atoms and/or positively charged ions each of which contains $N_e$ bound 
electrons ($N_e \ge 2$) and one central atomic nucleus with the electrical charge $Q$  
\begin{eqnarray} 
 &&d\sigma = \Bigl(\frac{8 \pi \alpha a^{2}_{0}}{9 \omega}\Bigr) \; \frac{4^{b} \; \nu^{2 b + 2} (1 + 
 \nu^{2}) \Gamma^{2}(b + 2)}{(\nu^{2} + b^{2})^{b+3} (1 - \exp(-2 \pi \nu)) \Gamma(2 b + 1)} \; 
 \exp\Bigl[- 4 \nu \; {\rm arccot}\Bigl(\frac{\nu}{b}\Bigr)\Bigr] \Bigr| (b - 1) \times \nonumber \\
 && (\nu - \imath b) \; \; {}_2F_{1}(2 - \imath \nu; 2 - b; 4; \frac{\nu - \imath b}{\nu + \imath b} \Bigr) 
 - \nu \; (b + 2) \; {}_2F_{1}(2 - \imath \nu, 1 - b; 4; \frac{\nu - \imath b}{\nu + \imath b} \Bigr) 
 \Bigr|^{2} \; \label{dsigmaZpol} \\
 && ({\bf n} \cdot {\bf e})^{2} do \; , \; \nonumber 
\end{eqnarray} 
where the incident beam of light is completely polarized. For natural light we obtain the following formula 
\begin{eqnarray} 
 &&d\sigma = \Bigl(\frac{4 \pi \alpha a^{2}_{0}}{9 \omega}\Bigr) \; \frac{4^{b} \; \nu^{2 b + 2} (1 + \nu^{2}) 
 \Gamma^{2}(b + 2)}{(\nu^{2} + b^{2})^{b+3} (1 - \exp(-2 \pi \nu)) \Gamma(2 b + 1)} \; \exp\Bigl[- 4 \nu \; 
 {\rm arccot} \Bigl(\frac{\nu}{b}\Bigr)\Bigr] \Bigr| (b - 1) \times \nonumber \\
 && (\nu - \imath b) \; \; {}_2F_{1}(2 - \imath \nu; 2 - b; 4; \frac{\nu - \imath b}{\nu + \imath b} \Bigr) - 
 \nu \; (b + 2) \; {}_2F_{1}(2 - \imath \nu, 1 - b; 4; \frac{\nu - \imath b}{\nu + \imath b} \Bigr) \Bigr|^{2} 
 \times \; \label{dsigmaZu} \\
 && ({\bf n}_l \times {\bf n})^{2} do \; \; . \; \; \nonumber 
\end{eqnarray}
 
Finally, the corresponding formula for the total cross section of photoionization of the $(Q, N_e)-$atomic system, 
where $N_e \ge 2$, takes the form
\begin{eqnarray} 
 && \sigma = \Bigl(\frac{32 \pi^{2} \alpha a^{2}_{0}}{27 \omega}\Bigr) \; \frac{4^{b} \; \nu^{2 b + 2} (1 + 
 \nu^{2}) \Gamma^{2}(b + 2)}{(\nu^{2} + b^{2})^{b+3} (1 - \exp(-2 \pi \nu)) \Gamma(2 b + 1)} \; \exp\Bigl[- 4 \nu 
 \; {\rm arccot}\Bigl(\frac{\nu}{b}\Bigr)\Bigr] \times \; \label{sigmaZ} \\
 && \Bigr| (b - 1) (\nu - \imath b) \; \; {}_2F_{1}(2 - \imath \nu; 2 - b; 4; \frac{\nu - \imath b}{\nu + \imath 
 b} \Bigr) - \nu \; (b + 2) \; {}_2F_{1}(2 - \imath \nu, 1 - b; 4; \frac{\nu - \imath b}{\nu + \imath b} \Bigr) 
 \Bigr|^{2} \; . \; \nonumber
\end{eqnarray} 
Note that the parameter $b$ in these formulas is a real number, which is usually bounded between 0 and 2, i.e., $0 
< b < 2$. This means that the hypergeometric functions in Eqs.(\ref{rint1}) and (\ref{rint2}) can be determined only 
numerically (two exceptional cases when $b = 1$ and $b = 0$ are considered in the next two Sections). Recently, a 
number of fast, reliable and numerically stable algorithms have been developed and tested for accurate calculations 
of the hypergeometric functions. Our final formula can be simplified even further, if one applies the following 
relation (see, e.g., Eq.15.2.3 in \cite{AS}) between two hypergeometric functions which are included in our 
Eq.(\ref{I2rdmod}):
\begin{eqnarray} 
 & &{}_2F_{1}(2 - \imath \nu, 2 - b; 4; \frac{\nu - \imath b}{\nu + \imath b} \Bigr) = 
  {}_2F_{1}(2 - \imath \nu, 1 - b; 4; \frac{\nu - \imath b}{\nu + \imath b} \Bigr) \nonumber \\
 &+& \frac{1}{1 - b} \Bigl(\frac{\nu - \imath b}{\nu + \imath b}\Bigr) \frac{d }{d z} \Bigl[ {}_2F_{1}(2 
 - \imath \nu, 1 - b; 4; z) \Bigr] \; \; , \; \; \label{15.2.3}
\end{eqnarray}
where $z = \frac{\nu - \imath b}{\nu + \imath b}$. This formula allows one to operate with one hypergeometric 
function only. 

All formulas derived and presented in this Section can directly be used to determine the both differential and total 
cross sections of photodetachment of the outer-most electrons in few- and many-electron neutral atoms and positively 
charged ions which contains $N_e$ bound electron, where $N_e \ge 2$. Note also that the total cross section of 
photoionization of the $(Q, N_e)-$atomic system, Eq.(\ref{sigmaZ}) can be re-written in a slightly different form 
$\sigma(\omega) = \sigma(\omega; I, b)$ which shows the explicit dependence of the photoionization cross section upon 
the cyclic frequency of incident light $\omega$. Such explicit formulas are very popular among people which conduct 
experiments and those theorists who calculate the convolution of various energy spectra, e.g., thermal spectra, and 
photodetachment cross sections. To achieve this goal one needs to replace variables in Eq.(\ref{sigmaZ}) by using the 
following (equivalent) expression for $\nu$ written in terms of $I, \omega$ and $b$: $\nu = b \sqrt{\frac{I}{\omega 
- I}}$. Finally, one finds the following formula for the total cross section $\sigma = \sigma(\omega, I, b)$ of 
photodetachment of the outer-most electrons in few-electron neutral atoms and positively charged ions
\begin{eqnarray} 
 \sigma&=&\Bigl(\frac{32 \pi^{2} \alpha a^{2}_{0}}{27}\Bigr) \; \frac{4^{b} \; I^{b + 2} \; \Bigl[1 + (b^{2} - 1) 
 \frac{I}{\omega}\Bigr] \; \; \Gamma^{2}(b + 3)}{b^{4} \omega^{b+3} \; [1 - \exp(-2 \pi b \sqrt{\frac{I}{\omega - 
 I}}\Bigr)\Bigr] \; \Gamma(2 b + 1)} \; \exp\Bigl[- 4 b \sqrt{\frac{I}{\omega - I}} \; {\rm 
 arccot}\Bigl(\sqrt{\frac{I}{\omega - I}}\Bigr)\Bigr] \nonumber \\
 &\times&\Bigl| \frac{b - 1}{b + 2} \; \Bigl(1 - \imath b \sqrt{\frac{\omega - I}{I}}\Bigr) \; \; {}_2F_{1}(2 
 - \imath b \sqrt{\frac{I}{\omega - I}}, 2 - b; 4; \frac{\sqrt{I} - \imath \sqrt{\omega - I}}{\sqrt{I} + \imath 
 \sqrt{\omega - I}} \Bigr) \nonumber \\
 &-& {}_2F_{1}(2 - \imath b \sqrt{\frac{I}{\omega - I}}, 1 - b; 4; \frac{\sqrt{I} - \imath \sqrt{\omega 
 - I}}{\sqrt{I} + \imath \sqrt{\omega - I}} \Bigr) \Bigr|^{2} \; . \label{sigmaZ1}
\end{eqnarray} 
This formula is one of the main results of this study. For one-electron atomic systems, when $b = 1$, the last 
formula, Eq.(\ref{sigmaZ1}), takes a familiar form: 
\begin{eqnarray} 
 \sigma&=&\Bigl(\frac{32 \pi^{2} \alpha a^{2}_{0}}{27}\Bigr) \; \frac{4 \; I^{3} \; \; \Gamma^{2}(4)}{\omega^{4} \; 
 [1 - \exp(-2 \pi \sqrt{\frac{I}{\omega - I}}\Bigr)\Bigr] \; \Gamma(3)} \; \exp\Bigl[- 4 \sqrt{\frac{I}{\omega - 
 I}} \; {\rm arccot} \; \Bigl(\sqrt{\frac{I}{\omega - I}}\Bigr)\Bigr] \nonumber \\ 
 &=& \frac{256 \pi^{2} \alpha a^{2}_{0}}{3} \; \frac{I^{3}}{\omega^{4} \; [1 - \exp(-2 \pi \sqrt{\frac{I}{\omega 
  - I}}\Bigr)\Bigr]} \; \exp\Bigl[- 4 \sqrt{\frac{I}{\omega - I}} \; {\rm arccot} \; \Bigl(\sqrt{\frac{I}{\omega 
  - I}}\Bigr)\Bigr] \; , \; \label{FFF}
\end{eqnarray} 
which exactly coincides with another our formula, Eq.(\ref{sigmaQ}), derived below (Stobbe's formula). Indeed, for 
one-electron atomic systems we have $b = 1$ and $I = \frac{Q^{2}}{2}$, and therefore, $\frac{I^{3}}{\omega^{4}} = 
\Bigl(\frac{I}{\omega}\Bigr)^{4} \frac{2}{Q^{2}}$. This is the first known derivation of the formula, Eq.(\ref{sigmaQ}), 
(or Stobbe's formula) for one-electron atom and ions from our much more general and universal expression, Eq.(\ref{FFF}), 
which is also applicable for arbitrary few- and many-electron atoms and positively charged ions. This problem was earlier 
considered by Hans Bethe (see \cite{BS} and earlier references therein), but he could not derive any closed analytical 
expression for the photoionization cross section similar to our formula, Eq.(\ref{FFF}) for the few- and many-electron 
atoms/ions. 
   
The both formulas, Eqs.(\ref{dsigmaZu}) and (\ref{sigmaZ1}), can be re-written in slightly different forms, if we 
introduce the universal photoionization function $F_{+}(b, x) = F_{+}(b, \frac{I}{\omega})$, which is defined for 
few- and many-electron atoms and positively charged atomic ions by the equation:
\begin{eqnarray}
 &&F_{+}(b, x) = \frac{4^b \Gamma(b + 3)}{b^{4} \Gamma(2 b + 1)} \; \frac{x^{b+3} [1 + (b^2 - 1) x]}{1 - 
 \exp(-2 \pi b y)} \times \nonumber \\
 &&\Bigl| \frac{b - 1}{b + 2} \; \Bigl(1 - \imath \frac{b}{y}\Bigr) \; \; {}_2F_{1}(2 - \imath \frac{b}{y}, 2 
  - b; 4; \frac{y - \imath}{y + \imath}\Bigr) - {}_2F_{1}(2 - \imath \frac{b}{y}, 1 - b; 4; \frac{y - \imath}{y 
  + \imath}\Bigr) \Bigr|^{2} \; \; , \; \label{F+}
\end{eqnarray} 
where $x = \frac{I}{\omega}, y = \sqrt{\frac{x}{1 - x}}$ and $b = \frac{Q - N_e + 1}{\sqrt{2 I}}$. By using this 
universal photoionization function $F_{+}(b, x)$ we can write the following, compact expressions for the differential 
and total cross sections    
\begin{eqnarray} 
  \frac{d \sigma}{do} = \Bigl(\frac{4 \pi \alpha a^{2}_{0}}{9}\Bigr) F_{+}\Bigl(b, \frac{I}{\omega}\Bigr) \; 
  ({\bf n}_l \times {\bf n})^{2} \; \; \; \; {\rm and} \; \; \; \; \sigma(\omega) = \Bigl(\frac{32 \pi^{2} \alpha 
  a^{2}_{0}}{27}\Bigr) F_{+}\Bigl(b, \frac{I}{\omega}\Bigr) \; \; \; . \; \label{F+A} 
\end{eqnarray} 

\section{One-electron atoms and ions}

Photoionization of one-electron atomic systems is significantly simpler than photodetachment of the outer-most electrons 
in a few- and many-electron atomic systems considered above. Indeed, in this case we have the atomic ionization potential 
$I$ which depends upon the nuclear charge $Q$ only, i.e., $I = I(Q)$. Furthermore, for the ground (bound) state in 
one-electron atoms and ions we always have $2 I = Q^{2}$, and therefore, in the both formulas, Eqs.(\ref{rint1}) and 
Eq.(\ref{rint2}) the parameter $b = 1$ and the first term in Eq.(\ref{diff1}) equals zero identically. Also, for $b 
= 1$ the hypergeometric functions ${}_2F_{1}(2 - \imath \nu, 0; 4; z)$, which is included in the second term, equals 
unity. This means that for one-electron atoms/ions (or for $b = 1$) we can express the both differential and total cross 
sections of photoionization in terms of elementary functions only. The normalization constant of the incident wave 
function equals $C(b;B) = C(1;Q) = \frac{Q \sqrt{Q}}{\sqrt{\pi}}$ and the two parameters $\nu$ and $\frac{\nu}{b}$ are 
now identical. This means that for any one-electron atom and/or ion the long-range asymptotics of its actual wave 
function always coincides with the wave function itself, and for ground states it is also coincides with the formula, 
Eq.(\ref{rad3}). In other words, by applying our method to the ground states in one-electron atomic systems we obtain 
the exact solution which correctly describes the non-relativistic photoionization (or photoeffect).  

First, we note that for one-electron atoms and ions the explicit formula for the auxiliary radial integral $I_{rd}$ in 
Eq.(\ref{ME}) (or $I^{(2)}_{rd}$ in Eq.(\ref{diff1})) takes the form    
\begin{eqnarray} 
 & &I_{rd} = - p Q^{2} \frac{Q \sqrt{Q}}{\sqrt \pi} \sqrt{\frac{2 \pi (1 + \nu^{2})}{\nu (1 - \exp(-2 \pi \nu))}} \; 
 \int_{0}^{+\infty} r^{(4 - 1)} \exp(- Q r -\imath p r) {}_1F_{1}(2 + \imath \nu; 4; 2 \imath p r) dr \nonumber \\
 &=& - p Q^{3} \sqrt{Q} \sqrt{\frac{2 (1 + \nu^{2})}{\nu (1 - \exp(-2 \pi \nu))}} \; \frac{\Gamma(4)}{(Q + 
 \imath p)^{4}} \; \; {}_2F_{1}(2 + \imath \nu; 4; 4; \frac{2 \imath p}{Q + \imath p} \Bigr) \; , \; \label{rint2a}
\end{eqnarray} 
where $\nu = \frac{Q}{p}$ and $\Gamma(4) = 3 \cdot 2 \cdot 1 = 6$. The hypergeometric function in the last equation can 
be transformed to the form 
\begin{eqnarray}
 {}_2F_{1}(2 + \imath \nu; 4; 4; \frac{2 \imath p}{Q + \imath p} \Bigr) = \Bigl(1 - \frac{2 \imath p}{Q + \imath p} 
 \Bigr)^{4 - 4 - 2 - \imath \nu} {}_2F_{1}(2 - \imath \nu; 0; 4; \frac{2 \imath p}{Q + \imath p} \Bigr) = \Bigl(\frac{Q 
 + \imath p}{Q - \imath p} \Bigr)^{2 + \imath \nu} \; , \label{hgf}
\end{eqnarray} 
where we applied the formula ${}_2F_{1}(\alpha, \beta; \gamma; z) = (1 - z)^{\gamma - \alpha - \beta} {}_2F_{1}(\gamma 
- \alpha, \gamma - \beta; \gamma; z)$ (see, Eq.(9.131) in \cite{GR}). Another way to obtain the same formula, 
Eq.(\ref{hgf}), is to apply the following expression for the integral in Eq.(\ref{rint2a}) with the confluent 
hypergeometric function(s) (see, e.g., \cite{LLQ} and \cite{GR})
\begin{eqnarray}
 \int_{0}^{+\infty} \exp(- \lambda z) z^{\gamma - 1} {}_1F_{1}(\alpha; \gamma; b z) dz = 
 \frac{\Gamma(\gamma)}{\lambda^{\gamma}} \; \Bigl(\frac{\lambda}{\lambda - b}\Bigr)^{\alpha} \; \; . \; 
\end{eqnarray}
Finally, we obtain 
\begin{eqnarray} 
 I_{rd} &=& - 6 Q^{3} \sqrt{Q} \sqrt{\frac{2 (1 + \nu^{2})}{\nu (1 - \exp(-2 \pi \nu))}} \; \frac{1}{(Q^{2} +  
 p^{2})^{2}} \; \Bigl(\frac{\nu + \imath}{\nu - \imath}\Bigr)^{\imath \nu} \nonumber \\
 &=& - \frac{6 Q^{3} \sqrt{Q}}{p^{4}} \sqrt{\frac{2}{\nu (1 - \exp(-2 \pi \nu)) \; (1 + \nu^{2})^{3}}} \; 
 \; \frac{\exp( -2 \nu \; {\rm arccot} \; \nu)}{(1 + \nu^{2})^{2}} \; \; . \; \label{1Ird}
\end{eqnarray} 

By using the notation $\nu = \frac{Q}{p}$ and multiplying the radial integral $I_{rd}$ by the factor $\frac{4 
\pi}{3} ({\bf n} \cdot {\bf e})$ we derive the following formula 
\begin{eqnarray} 
 \frac{4 \pi}{3} ({\bf n} \cdot {\bf e}) I_{rd} = - \frac{8 \pi}{\sqrt{p}} \sqrt{\frac{2 \nu^{6}}{[1 - \exp(-2 
 \pi \nu)] (1 + \nu^{2})^{3}}} \; \exp( -2 \nu \; {\rm arccot} \; \nu) ({\bf n} \cdot {\bf e}) \; . \; 
 \label{1Irda}
\end{eqnarray} 
Now, the explicit formula for the $\frac{p \alpha a^{2}_{0}}{2 \pi \omega} \mid I_{rd} \mid^{2}$ factor in 
Eq.(\ref{crossau}) takes the form 
\begin{eqnarray} 
 \frac{p \alpha a^{2}_{0}}{2 \pi \omega} \mid I_{rd} \mid^{2} = \frac{64 \pi \alpha a^{2}_{0}}{\omega} 
 \Bigl(\frac{\nu^{2}}{1 + \nu^{2}}\Bigr)^{3} \; \frac{\exp( -4 \nu \; {\rm arccot} \; \nu)}{1 - 
 \exp(-2 \pi \nu)} \; \; . \; \label{MatrEl}
\end{eqnarray} 
The final formula for the differential cross sections of photoionization of one-electron atom/ion by a completely 
polarized light is written in the form 
\begin{eqnarray} 
 d\sigma &=& \frac{64 \pi \alpha a^{2}_{0}}{\omega} \Bigl(\frac{\nu^{2}}{1 + \nu^{2}}\Bigr)^{3} \; 
 \frac{\exp( -4 \nu \; {\rm arccot} \; \nu)}{1 - \exp(-2 \pi \nu)} ({\bf n} \cdot {\bf e})^{2} do \nonumber \\
 &=& \frac{128 \pi \alpha a^{2}_{0}}{Q^{2}} \Bigl(\frac{I}{\omega}\Bigr)^{4} \; \frac{\exp( -4 \nu \; 
 {\rm arccot} \; \nu)}{1 - \exp(-2 \pi \nu)} ({\bf n} \cdot {\bf e})^{2} do \; , \; \label{dsigmaQpol}
\end{eqnarray} 
where we also used the following relation $1 = \frac{Q^{2}}{Q^{2}} = 2 \Bigl(\frac{Q^{2}}{2}\Bigr) \frac{1}{Q^{2}} 
= \frac{2 I}{Q^{2}}$. For the natural (or unpolarized) light the differential cross section of photoionization is 
written in the form
\begin{eqnarray} 
 d\sigma =  \frac{64 \pi \alpha a^{2}_{0}}{Q^{2}} \Bigl(\frac{I}{\omega}\Bigr)^{4} \; \frac{\exp( -4 \nu \; 
 {\rm arccot} \; \nu)}{1 - \exp(-2 \pi \nu)} ({\bf n} \times {\bf n}_{l})^{2} do \; . \; \label{dsigmauQunpol} 
\end{eqnarray}
The total cross section of photoionization for one-electron atom/ion with the nuclear charge $Q$ is 
\begin{eqnarray}
 \sigma &=& 512 \pi^{2} \alpha \; \Bigl(\frac{a^{2}_{0}}{Q^{2}}\Bigr) \; \Bigl(\frac{I}{\omega}\Bigr)^{4} \; 
 \frac{\exp( -4 \nu \; {\rm arccot} \; \nu)}{1 - \exp(-2 \pi \nu)} \nonumber \\
 &=& 512 \pi^{2} \alpha \; \Bigl(\frac{a^{2}_{0}}{Q^{2}}\Bigr) \; \Bigl(\frac{I}{\omega}\Bigr)^{4} \; 
 \frac{\exp\Bigl( -4 \sqrt{\frac{I}{\omega - I}} \; {\rm arccot} \; \sqrt{\frac{I}{\omega - I}}\Bigr)}{1 - 
 \exp\Bigl(-2 \pi \sqrt{\frac{I}{\omega - I}}\Bigr)} \; \; . \; \label{sigmaQ} 
\end{eqnarray} 
This formula can also be re-written in the form 
\begin{eqnarray}
 \sigma = \frac{256 \pi^{2} \alpha a^{2}_{0}}{I} \; \Bigl[ \frac{x^{4} \; \exp( -4 \; y \; {\rm arccot} \; y)}{1 
 - \exp(- 2 \pi \; y)} \Bigr] = \frac{256 \pi^{2} \alpha a^{2}_{0}}{I} \; F_{Q}(x) \; , \; \label{SigmaQF} 
\end{eqnarray} 
where $x = \frac{I}{\omega}, y = \sqrt{\frac{x}{1 - x}}$ and $F_{Q}(x)$ is the universal photoionization function 
defined for one-electron atom(s) and positively charged ions in which the nuclear charge of the central nucleus 
equals $Q$. 

Our formulas, Eqs.(\ref{dsigmauQunpol}), (\ref{sigmaQ}) and (\ref{SigmaQF}), exactly coincide with the formula 
obtained by Stobbe in \cite{Stob} and with the analogous formulas derived in \$ 56 from \cite{BLP}. Note also 
that the last formulas, Eqs.(\ref{dsigmaQpol}) - (\ref{SigmaQF}), have directly been derived from our formula, 
Eq.(\ref{sigmaZ1}), obtained in the previous Section for $b = 1$, but here we wanted to derive and check them 
by using an independent approach. As mentioned above for one-electron atoms and ions its ionization potential 
$I$ is the explicit (and simple) function of the nuclear charge $Q$ only. Generalization of our formulas to 
photodetachment of the outer-most electron from the excited atomic states is also simple and transparent, but 
it cannot be done directly with the use of DFT theory, since Eq.(\ref{rad3}) does not work for the excited 
states. Derivation of the explicit formulas for photoionization cross sections of the excited one-electron 
atoms and ions also requires additional explanations, extra notations and extensive analytical work.  

\section{Negatively charged ions} 

Investigation of the non-relativistic photoeffect in the negatively charged ions is reduced to the analysis of 
photodetachment of the outer-most electron in similar atomic systems. Note that photodetachment of the negatively 
charged atomic ions is of great interest in numerous applications. In general, there is a fundamental difference 
in photodetachment of the negatively charged ions and photoionization of the positively charged atomic ions and 
neutral atoms. In particular, for all negatively charged ions the electrical charge of the final atom $Z = Q - 
N_e + 1$ equals zero identically. Therefore, in this case we cannot introduce the parameter $\nu = \frac{Z}{p}$, 
which was very helpful in the two previous Sections. This means that all our formulas, derived for the 
photodetachment cross sections (see below), contain only the momentum of photo-electron $p$ and ionization 
potential $I$, or parameter $B = \sqrt{2 I}$. These two variables $p$ and $I$ (or $B$) are crucial for theoretical 
analysis of the non-relativistic photoeffect in the negatively charged ions, or their photodetachment. Such a 
photodetachment of the negatively charged, two-electron H$^{-}$ ion is of great interest for our understanding of 
all details in actual visible and infrared spectra of many stars, including our Sun. Photodetachment of the 
four-electron negatively charged Li$^{-}$ ion plays some role in developing of very compact and reliable 
photo-elements and recharged batteries. Therefore, it is important to produce some universal formula for the 
photodetachment cross sections of the negatively charged ions.    

For the negatively charged atomic ions the derivative of the radial wave function of the initial state, 
Eq.(\ref{rad3}), is written in the form 
\begin{eqnarray} 
 \frac{d}{d r} \Bigl[ \frac{C}{r} \exp(- B r) \Bigr] = - C r^{-2} \exp(- B r) - C B r^{-1} \exp(- B r) \; , 
 \; \label{diff2} 
\end{eqnarray} 
where $C$ is the normalization constant which equals $\sqrt[4]{\frac{I}{2 \pi^{2}}}$ as follows from Eq.(\ref{norm}) 
for $b = 0$. Therefore, the formula for our auxiliary radial integral $I_{rd}$, Eq.(\ref{aux}), will also include two 
different terms, i.e., $I_{rd} = 3 \Bigl( J^{(1)}_{rd} + J^{(2)}_{rd} \Bigr)$, where 
\begin{eqnarray} 
 & &J^{(1)}_{rd} = C \sqrt{\frac{\pi}{2 p}} \int_{0}^{+\infty} dr J_{\frac32}(p r) r^{\frac12 - 1} \exp(- B r) 
 = C \sqrt{\frac{\pi}{2 p}} \Bigl(\frac{p}{2}\Bigr)^{\frac32} \frac{\Gamma(2)}{\Gamma\Bigl(\frac52\Bigr) (B^{2} 
 + p^{2})} \times \; \; \label{J1a} \\
 & &{}_2F_{1}\Bigl(1, 1; \frac52; \frac{p^{2}}{B^{2} + p^{2}} \Bigr) = \frac{C p}{3 (B^{2} + p^{2})} \; \; 
 {}_2F_{1}\Bigl(\frac32, \frac32; \frac52; \frac{p^{2}}{B^{2} + p^{2}} \Bigr) \; . \; \nonumber
\end{eqnarray} 
The hypergeometric function in the last equation can be reduced to some combination of elementary functions. To 
show this explicitly let us apply the following formula 
\begin{eqnarray}
 \frac{d}{d z} \; \; \Bigl[ {}_2F_{1}(\alpha, \beta; \gamma; z) \Bigr] = \frac{\alpha \beta}{\gamma} \; \; 
 {}_2F_{1}(\alpha + 1, \beta + 1; \gamma + 1; z) \; \; , \; \label{HHG1}
\end{eqnarray} 
where in our case $\alpha = \frac12, \beta = \frac12$ and $\gamma = \frac32$. For these values of $\alpha, 
\beta$ and $\gamma$ the last formula takes the form 
\begin{eqnarray}
 \frac{d}{d z} \; \; \Bigl[ {}_2F_{1}\Bigl( \frac12, \frac12; \frac32; z ) \Bigr] = \frac16 \; \; {}_2F_{1}\Bigl( 
 \frac32, \frac32;  \frac52; z \Bigr) \; , \; \label{HHG2}
\end{eqnarray} 
where the argument $z$ varies between zero and unity, i.e., $0 \le z < 1$. In our case this is true, since $z = 
\frac{p^{2}}{B^{2} + p^{2}}$. Now, we can write 
\begin{eqnarray}
 {}_2F_{1}\Bigl( \frac32, \frac32; \frac52; z\Bigr) = 6 \frac{d}{d z} \Bigl[ {}_2F_{1}\Bigl(\frac12; \frac12; 
 \frac32; z) \Bigr] = 6 \frac{d}{d z} \Bigl( \frac{\arcsin \sqrt{z}}{\sqrt{z}} \Bigr) = 3 \frac{\sqrt{z} - 
 \sqrt{1 - z} \; \arcsin \sqrt{z}}{z \sqrt{z (1 - z)}} \; , \;  \label{HHG3}
\end{eqnarray} 
where we used the formula Eq.(15.1.6) from \cite{AS} for the ${}_2F_{1}(\frac12; \frac12; \frac32; z)$ hypergeometric 
function, i.e., ${}_2F_{1}(\frac12; \frac12; \frac32; z) = \frac{\arcsin \sqrt{z}}{\sqrt{z}}$. The analytical formula, 
Eq.(\ref{HHG3}), derived for this ${}_2F_{1}\Bigl( \frac32, \frac32; \frac52; z \Bigr)$ function is our original 
result which cannot be found directly neither in \cite{GR}, nor in \cite{AS}. In our case $z = \frac{p^{2}}{B^{2} + 
p^{2}}, \sqrt{1 - z} = \frac{B}{\sqrt{B^{2} + p^{2}}}$ and $\sqrt{z} = \frac{p}{\sqrt{B^{2} + p^{2}}}$ and the final 
formula for the $J^{(1)}_{rd}$ integral takes the form 
\begin{eqnarray} 
 J^{(1)}_{rd} = \frac{C}{\sqrt{B^{2} + p^{2}}} \; \; \frac{\sqrt{z} - \sqrt{1 - z} \arcsin\sqrt{z}}{z \sqrt{1 
 - z}} \; \; . \; \label{J1rd}
\end{eqnarray} 
The expression in the right-hand side of this equation is not singular when $z \rightarrow 0$ (or $p 
\rightarrow 0$), since
\begin{eqnarray} 
 \lim_{z \rightarrow 0} \frac{\sqrt{z} - \sqrt{1 - z} \arcsin\sqrt{z}}{z \sqrt{1 - z}} = \frac16 \lim_{z 
 \rightarrow 0} \sqrt{z} = 0 \; \; . \; \nonumber 
\end{eqnarray} 

Analogous formula for the second radial integral $J^{(2)}_{rd}$ is 
\begin{eqnarray} 
 & &J^{(2)}_{rd} = C \; B \sqrt{\frac{\pi}{2 p}} \int_{0}^{+\infty} dr J_{\frac32}(p r) r^{\frac32 - 1} \exp(- B r) 
 = \frac{C}{\sqrt{2 p}} \Bigl(\frac{p}{2}\Bigr)^{\frac32} \frac{B \; \Gamma(3)}{\Gamma\Bigl(\frac52\Bigr) (B^{2} + 
 p^{2})^{\frac32}} \times \; \; \nonumber \\
 & &{}_2F_{1}(\frac32; \frac12; \frac52; \frac{p^{2}}{B^{2} + p^{2}} \Bigr) = \frac{2 \; C \; B \; p}{3 (B^{2} + 
 p^{2})^{\frac32}} \; \; {}_2F_{1}(\frac32; \frac12; \frac52; \frac{p^{2}}{B^{2} + p^{2}} \Bigr) \; . \; \label{J2a}
\end{eqnarray}
It is possible to obtain the explicit expression for the ${}_2F_{1}\Bigl(\frac32; \frac12; \frac52; z)$ hypergeometric 
function in terms of some elementary functions. For this purpose we need to use the known analytical formula for the 
${}_2F_{1}\Bigl( \frac12, \frac12; \frac32; z )$ function (which equals $\frac{\arcsin \sqrt{z}}{\sqrt{z}}$, see, 
Eq.(\ref{HHG3})) and apply the formula Eq.(15.2.7) from \cite{AS} for $n = 1$ which takes the form 
\begin{eqnarray}
 \frac{d }{d z} \Bigl[ (1 - z)^{a} {}_{2}F_{1}( a, b; c; z) \Bigr] = - \frac{a (c - b)}{c} (1 - z)^{a - 1} 
 {}_{2}F_{1}( a + 1, b; c + 1; z) \; , \; \label{ASint1}  
\end{eqnarray} 
where $a = \frac12, b = \frac12$ and $c = \frac32$. Now, for the ${}_2F_{1}\Bigl( \frac12, \frac12; \frac32; z )$ 
function one finds 
\begin{eqnarray}
 \frac{d }{d z} \Bigl[ (1 - z)^{\frac12} {}_{2}F_{1}\Bigl( \frac12, \frac12; \frac32; z) \Bigr] = - \frac{1}{3} 
 (1 - z)^{-\frac12} \; {}_{2}F_{1}( \frac32, \frac12; \frac52; z) \; . \;  \label{ASint2}   
\end{eqnarray}
From this equation we derive
\begin{eqnarray}
 & & {}_{2}F_{1}\Bigl( \frac32, \frac12; \frac52; z) = -3 \sqrt{1 - z} \; \frac{d }{d z} \Bigl[ \sqrt{1 - z} 
 \Bigl(\frac{ \arcsin \sqrt{z}}{\sqrt{z}} \Bigr) \Bigr] = \frac32 \Bigl( \frac{\arcsin \sqrt{z}}{\sqrt{z}} \; 
 \nonumber \\
 &-& \frac{\sqrt{1 - z}}{z} + \frac{1 - z}{z \sqrt{z}} \arcsin\sqrt{z} \Bigr) = \frac{3}{2 \sqrt{z}} \Bigl( 
 \frac{\arcsin  \sqrt{z}}{z} - \frac{\sqrt{z (1 - z)}}{z} \Bigr) \; \; , \; \label{ASint25}  
\end{eqnarray}
where $z = \frac{p^{2}}{B^{2} + p^{2}} < 1$ and $z \ge 0$. This analytical formula for the ${}_2F_{1}\Bigl( 
\frac32, \frac12; \frac52; z \Bigr)$ function is another original result which cannot be found neither in 
\cite{GR}, nor in \cite{AS}. Note also that analytical formula for the integral in Eq.(\ref{J2a}) can also 
be derived as a partial derivative of the $J^{(1)}_{rd}$ in respect to the parameter $B$. Thus, the both 
auxiliary radial integrals $J^{(1)}_{rd}$ and $J^{(2)}_{rd}$ are expressed in terms of the elementary 
functions. In particular, the final formula for the $J^{(2)}_{rd}$ integral is
\begin{eqnarray} 
 J^{(2)}_{rd} = \frac{C}{\sqrt{B^{2} + p^{2}}} \; \sqrt{1 - z} \; \Bigl[ \frac{\arcsin \sqrt{z}}{z} - 
 \sqrt{\frac{1 - z}{z}} \Bigr]  \; , \; \label{J2rd} 
\end{eqnarray}
where $z = \frac{p^{2}}{B^{2} + p^{2}}, \sqrt{z} = \frac{p}{\sqrt{B^{2} + p^{2}}}$ and $\sqrt{1 - z} = 
\frac{B}{\sqrt{B^{2} + p^{2}}}$. Again, by using the formula, Eq.(1.641) from \cite{GR} for the $\arcsin x$ one 
can easily show that 
\begin{eqnarray} 
 \lim_{z \rightarrow 0} \Bigl(\frac{\arcsin \sqrt{z}}{z} - \sqrt{\frac{1 - z}{z}} \Bigr) = 0 \; , \; \nonumber 
\end{eqnarray} 
which means that our formula for the $J^{(2)}_{rd}$ integral is not singular at $z \rightarrow 0$ (or at $p 
\rightarrow 0$). Note that the formula Eq.(\ref{crossau}) for the differential cross section of photodetachment 
always contains an additional factor $p$ in its numerator. Therefore, such a cross-section for an arbitrary 
negatively charged ion always approaches zero when $p \rightarrow 0$. The same statement is true for the total 
cross sections of photodetachment of the negatively charged ions. In contrast with this, analogous cross sections 
(differential and total) of atomic systems considered in the two previous Sections approach (when $p \rightarrow 
0$) some final limits which are not equal zero. All these features of photoionization cross sections of the 
neutral atoms and positively charged ions are well known from numerous experiments and theoretical calculations 
(see, e.g., \cite{BS}, \cite{Star} and references therein).

Analytical computations of the total auxiliary $I_{rd} = 3 \Bigl( J^{(1)}_{rd} + J^{(2)}_{rd} \Bigr)$ integral 
and the both differential and total cross sections are simple and straightforward. The final formula, 
Eq.(\ref{sigma-aa}), for the differential cross section of photodetachment of the negatively charged atomic 
ions takes the form 
\begin{eqnarray}
 d\sigma = \Bigl(\frac{8 \pi \alpha a^{2}_{0} p}{9 \omega}\Bigr) \; ({\bf n} \cdot {\bf e})^{2} \mid I_{rd} 
 \mid^{2} do = \frac{8 \alpha a^{2}_{0}}{\omega} \; \Bigl[\frac{p B}{\omega (B^{2} + p^{2})}\Bigr] \; \mid 
 J^{(1)}_{rd} + J^{(2)}_{rd} \mid^{2} ({\bf n} \cdot {\bf e})^{2} do \label{sigma-polar}
\end{eqnarray}
for completely polarized light. Analogous formula for unpolarized light is  
\begin{eqnarray}
 d\sigma = \frac{4 \alpha a^{2}_{0}}{\omega} \; \sqrt{\frac{1 - z}{z}} \; \Bigl[(\sqrt{1 - z} - 1) 
 \frac{\arcsin(\sqrt{z})}{\sqrt{z}} + \frac{1}{\sqrt{1 - z}} - 1 + z \Bigr]^{2} ({\bf n}_{l} \times 
 {\bf n})^{2} do , \label{sigma-ang} 
\end{eqnarray}
where $z = \frac{p^{2}}{B^{2} + p^{2}}, \sqrt{z} = \frac{p}{\sqrt{B^{2} + p^{2}}}, \; \sqrt{1 - z} = 
\frac{B}{\sqrt{B^{2} + p^{2}}}$ and $B^{2} = 2 I$. The formula for the total cross section is written in 
the form
\begin{eqnarray}
 \sigma &=& \frac{32 \pi \alpha a^{2}_{0}}{3 \omega} \; \sqrt{\frac{1 - z}{z}} \; \Bigl[(\sqrt{1 - z} 
 - 1) \frac{\arcsin(\sqrt{z})}{\sqrt{z}} + \frac{1}{\sqrt{1 - z}} - 1 + z \Bigr]^{2} \; \; \; \nonumber \\
 &=& \frac{32 \pi \alpha a^{2}_{0}}{3 I} \; \sqrt{\frac{x}{1 - x}} \; \Bigl[ (\sqrt{x (1 - x)} - 1) 
  \frac{\arcsin\sqrt{1 - x}}{\sqrt{1 - x}} + 1 - x \sqrt{x} \Bigr]^{2} = \frac{32 \pi \alpha a^{2}_{0}}{3 
  I} \; F_{-}(x) \; \; \; \label{sigma-totl} 
\end{eqnarray} 
where $z = 1 - \frac{I}{\omega}, x = 1 - z = \frac{I}{\omega}$ and $0 \le x < 1$. The function $F_{-}(x) = 
F_{-}\Bigl(\frac{I}{\omega}\Bigr)$ defined in this equation is the universal photodetachment function which 
is applied to an arbitrary negatively charged ion. Derivation of the formulas for the differential and total 
cross sections of photodetachment of the negatively charged atomic ions is one of the main results of this 
study. The $F_{-}(x)$ function defined in Eq.(\ref{sigma-totl}) is the same for all negatively charged ions 
which means that the cross sections of photodetachment for all negatively charged ions are similar (in this 
sense) to each other. Briefly, this means that there is no principal difference between photodetachment 
cross sections of the H$^{-}$, Li$^{-}$ and O$^{-}$ ions. It can be shown that the $F_{-}(x)$ function has 
one maximum in the area of our interest, i.e., for $0 \le x < 1$. The both amplitude and location of this 
maximum on the $\omega$ axis depend upon the ionization potential $I$ only. The differential cross section 
of photodetachment of an arbitrary negatively charged ion is also represented in a simple analytical form 
with the universal $F_{-}(x) = F_{-}\Bigl(\frac{I}{\omega}\Bigr)$ function. 
  
The formulas presented in this Section allow one to describe (completely and accurately) photoeffect in the 
negatively charged atomic ions. Derivation of our formulas for the differential and total cross sections of 
photodetachment of the negatively charged atomic ions is one of the main results of this study. All these 
formulas contain only elementary functions, and this was a real scientific surprise. Note also that our 
formula, Eq.(\ref{sigma-totl}), for the $\sigma = \sigma(\omega, I)$ dependence allows one to check and 
mainly confirm some earlier predictions made by Chandrasekhar in his papers about photodetachment 
cross-section of the negatively charged H$^{-}$ ion \cite{Chan1}, \cite{Chan2} (see also discussion in 
Sect.74 of \cite{BS}, \cite{John} and \cite{FroF} and references therein). The formulas derived above can 
also be used to describe photodetachment of the weakly-bound deuterium nucleus \cite{BetPei}. However, H. 
Bethe could not produce the closed analytical expressions for the photodetachment cross sections of the 
H$^{-}$ ion (similar to our Eqs.(\ref{sigma-ang}) and (\ref{sigma-totl})).      

\section{Discussion and Conclusions}  

In this study, by using the methods of quantum electrodynamics, we have developed the closed analytical 
procedure to describe the photoelectric effect in few-electron atoms and positively charged ions, as well as 
in the negatively charged atomic ions. The electron density distributions in the incident atoms/ions have 
been taken from DFT \cite{Osten}. Based on this procedure we have studied the non-relativistic photoeffect in 
different atomic systems including the neutral atoms and positively charged atomic ions which contain $N_e$ 
bound electrons ($N_e \ge 2$) in an atom/ion with the nuclear charge $Q$. Photoionization of one-electron 
atoms/ions and photodetachment of the negatively charged atomic ions, where $N_e = Q + 1$ (or $Q = N_e - 1$), 
are also investigated. In each of these cases we have derived the closed analytical formulas for the both 
differential and total cross sections of photoionization and/or photodetachment (see, Eqs.(\ref{dsigmaZpol}), 
(\ref{sigmaZ}), (\ref{sigma-ang}) and (\ref{sigma-totl}) above) of the corresponding atoms and ions. Note that 
each of these formulas contains only a few basic parameters of the original problem, e.g., the cyclic frequency 
of incident light $\omega$, atomic ionization potential $I$, the total number of initially bound electrons $N_e$ 
and electrical charge of atomic nucleus $Q$. These and some other formulas are the main results of this study. 
For neutral atoms and positively charged atomic ions with $N_e \ge 2$ and for negatively charged atomic ions 
similar formulas have never been produced in earlier papers. Our procedure developed here allows one to determine 
(both analytically and numerically) the differential and total cross sections of thermal photoionization of 
arbitrary few- and many-electron atomic systems. Our results obtained for the neutral He atom and Li$^{+}$ ion 
agree very well with the results of previous numerical calculations \cite{Bha2013}, \cite{BellLi+} of these 
systems. Maximal deviations of our total cross sections of photoeffect for these two systems and similar cross 
sections obtained in \cite{Bha2013} and \cite{BellLi+} do not exceed 8 \% - 10 \%. 
 
Our formulas derived for few-electron atoms and positively charged ions have been tested in applications to 
one-electron atoms and ions. Note that our analytical expressions for the differential and total cross 
sections of one-electron atomic systems have been derived in the both ways, i.e., directly and as a limit 
(when $N_e \rightarrow 1$) of the formulas obtained for photoionization of neutral atoms and positively 
charged ions. Remarkably, but all these formulas coincide with each other and with the well known formulas 
obtained earlier in \cite{Stob} and \cite{BLP}. Also, our analytical formulas derived for the photodetachment 
cross sections (differential and total) of the negatively charged ions are original and include only 
elementary functions. None of these formulas has ever been produced in earlier studies. By using our formulas 
we have determined the differential and total cross sections of the negatively charged H$^{-}$ ion which are 
in good agreement with fundamental calculations performed in \cite{AjCh} (see, also \cite{Bha2013} and 
references therein). For the negatively charged ions our differential and total cross sections for the 
negatively charged ions deviate from the results of numerical computations \cite{AjCh} does not exceed 18 \%. 

Thus, the quantitative agreement of our formulas with the known computational results for differential and total 
cross sections should be recognized as good, or even very good. At the same time, the qualitative agreement 
between our formulas and many results of experimental and theoretical investigations of analogous to atomic 
systems is absolute. This is true for all three classes of few-electron atomic systems, i.e., for all  
few-electron neutral atoms and ions, for the negatively charged ions and for one-electron atoms and ions. 
Currently, similar cross sections are determined either in numerical computations, or experimentally. However, 
from the  columns of numbers with many digits in each it is very hard (even impossible) to derive the 
corresponding analytical formulas which produce these results. Furthermore, based on the numerical approaches 
only one will always miss something important and interesting in our description of physical phenomena. This 
transforms our formulas derived in this paper into a very useful tool which can be used for educational and 
methodological purposes. Moreover, by using linear combinations of our formulas for the differential and total 
cross sections one can develop a simple analytical procedure to approximate actual cross sections for arbitrary  
frequencies $\omega$ to very high accuracy. 

Briefly, in this study we have shown that the methods of quantum electrodynamics can successfully be applied to 
describe photoelectric effect in few-electron atoms and positively charged ions, as well as in the negatively 
charged atomic ions. Our differential and total cross sections of the photoionization and photodetachment can 
now be evaluated to very good numerical accuracy, which is, probably, the best numerical accuracy known for 
single-electron approximations. These advantages and our analytical formulas for all cross sections are the main 
results of our work in this direction. It is clear that these results cannot compete with the results of very 
costly, system-oriented computations performed recently for some of these atomic systems, but this was not our 
aim in this study.

\end{document}